\documentclass[Journal]{IEEEtran}

\usepackage{caption}
\usepackage{subcaption}

\usepackage{setspace}

\usepackage{clrscode}
\usepackage{pict2e}
\usepackage{amsmath}
\usepackage{amssymb}
\usepackage{graphicx}
\usepackage{epsfig}
\usepackage[bookmarks=false]{hyperref}
\usepackage{multirow} %introduce multiple rows
\usepackage{amsmath}  %introduce \tag {}
\usepackage{algorithm}
\usepackage{algorithmic}
\newcommand{\E}{\mathbb{E}}

\newtheorem{Cor}{Corollary}

\newtheorem{Prop}{Proposition}

\begin{document}
\title{Energy-Efficient Resource Allocation for SWIPT in Multiple Access Channels}
\author{%\authorblockN{
Tewodros~A.~Zewde~and~M.~Cenk~Gursoy,~%}
\\
%\authorblockA{
Wireless Communication and Networking Lab\\
Department of Electrical Engineering and Computer Science\\
Syracuse University, Syracuse, NY 13244\\
tazewde@syr.edu, mcgursoy@syr.edu
}%}
\maketitle
\vspace{-0.8in}
\begin{abstract}
In this paper, we study optimal resource allocation strategies for simultaneous information and power transfer (SWIPT) focusing on the system energy efficiency. We consider two-user multiple access channels in which energy harvesting (EH) and information decoding (ID) nodes are spatially separated. We formulate optimization problems that maximize system energy efficiency while taking harvested energy constraints into account. These are concave-linear fractional problems, and hence Karush-Kuhn-Tucker (KKT) conditions are necessary and sufficient to obtain globally optimal solution. Solving these optimization problems, we provide analytical expressions for optimal transmit power allocation among the source nodes, and identify the corresponding energy efficiency. We confirm the theoretical analysis via numerical results. Furthermore, we also characterize the effect of circuit power consumption on the system's efficiency as the harvested energy demand varies. 
\end{abstract}
\thispagestyle{empty}
%\begin{keywords}
%Dual-power-splitting, multiple-access channels, relaying, spatial channel correlation, throughput, energy harvesting
%\end{keywords}
\section{Introduction}
\IEEEPARstart{S}{imultaneous} wireless information and
power transfer (SWIPT) aims to jointly transfer data as well as power to a destination where information-decoding (ID) and energy-harvesting (EH) operations can be carried out concurrently. The feasibility and implementation of wireless power transfer along with the data communication have been intensively studied in recent years. Numerous publications on simultaneous information and power transfer in the literature provide concrete theoretical frameworks and promising numerical results.\\
\indent Optimal parameters and characteristic regions designed for information transfer alone cannot be the same when power transfer is incorporated. Hence, it is essential to study the tradeoff characteristics and performance boundaries of SWIPT. Most of the above-mentioned works investigate the impact of harvested energy constraint on the achievable data rate, and they explicitly show this using the rate-energy tradeoff regions. However, efficient utilization of available energy to transfer every bit of information is an important and a compelling performance parameter \cite{Oba}. Several recent studies have addressed this key parameter in the presence of SWIPT under various settings. The authors in \cite{Robert} study energy efficiency (EE) of multiple users that operate using OFDMA scheme where each user has co-located ID and EH circuitries. In this work, harvested energy from a dedicated information-bearing signal and other nearby unknown interference is deducted from the consumption when evaluating the total number of bits successfully transfered per net consumed energy. A related work with similar EE definition is presented in reference \cite{Zhang} considering multiple-input single-output (MISO) models. Meanwhile, the authors in \cite{Xiong} have used the conventional EE definition, i.e., achievable data rate per total consumed energy, while studying energy-efficient OFDMA systems with non-overlapping uplink and downlink operations intervals. Unlike these approaches, the potential capacity produced by the transferred energy is added to the system capacity in reference \cite{Sun} assuming that the harvested energy primarily contributes to future information transfer. In all these definitions, EE metrics measure successfully transfered bits of information per consumed energy. On the other hand, the authors in \cite{Leng} introduce a new approach to evaluate the EE of SWIPT systems. They consider the EE of information transfer and EE of energy transfer separately, where the latter is defined as the amount of energy transfered per total consumption.\\
\indent All these studies provide interesting observations and new insights. However, only limited attention has been paid to energy-efficient resource allocation for multiple users communicating through multiple access channels as noted in \cite{Joung} \cite{Cheng} and \cite{Yu}. Therefore, optimal transmission strategies and closed-form expressions that explicitly characterize the corresponding optimal system energy efficiency in fading multiple access channels (MACs) remains unknown. Hence, with this motivation, we study optimal resource allocation among multiple users communicating through a MAC in the presence of SWIPT. More specifically, we investigate optimal resource allocation for two-user MAC maximizing the instantaneous energy efficiencies. We provide analytical expressions for the optimal transmission power and system efficiency. Using mathematical tools, we explicitly characterize the impact of harvested energy demand on the optimal EE and other critical parameters such as transmit power level. Since formulated optimizations are concave-linear fractional problems, we identify the Karush-Kuhn-Tucker (KKT) conditions to determine optimal solutions for every scenario. Finally, we provide detailed numerical results in order to confirm the theoretical analysis.
\section{System Model and Preliminaries}
\subsection{Model}
\begin{figure}[!ht]
	\centering
	\includegraphics[width=0.3\textwidth]{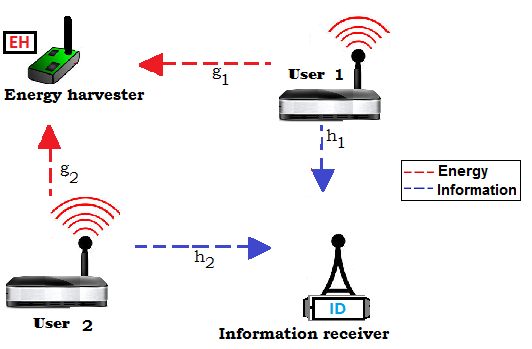}
	\caption{SWIPT in separated architecture}
	\label{Fig.ntk-b}
\end{figure}

We consider an uplink operation in which multiple users simultaneously send information-bearing signals on the same frequency band to a common receiver through a multiple access channel. The transmitted signal from node $i$ in the $k^{th}$ symbol duration is denoted by $X_i[k]\sim\mathcal{N}\big(0,P_i[k]\big)$, and its power is upper bounded as $\E\big\{|X_i[k]|^2\big\}= P_i[k] \leq P_i^{mx}$. The receiver decodes each user's information based on successive interference cancellation, and on this regard, we consider fixed decoding order. Meanwhile, these signals may also be used to transfer power to an energy harvesting (EH) node that can be either co-located or separated from an information decoding (ID) node. Thus, we study resource allocation among the transmitting nodes where the receiver carries out decoding and/or harvesting operations as shown in Fig. \ref{Fig.ntk-b}. We analyze system's energy efficiency assuming that each user targets wireless information and power transfer based on the spatial allocation of ID and EH components.
\\
\indent In regard to the wireless channel, we assume that the link between any user and the receiver experiences frequency-flat fading, and the corresponding channel power gain follows an exponential distribution. The complex fading coefficient for the channel between $i^{th}$ user and a receiver is denoted by $r_i$, and $h_i=|r_i|^2$ with the marginal probability density function $f_h(h_i)=\frac{1}{a}e^{(-\frac{h_i}{a})}$. Furthermore, each transmitting as well as receiving node is equipped with a single antenna, and channel state information is available at both ends. Therefore, the received signal at the destination in the $k^{th}$ symbol duration is given as
\begin{equation}
	Y[k]=\sum_{i=1}^2r_iX_i[k]+N[k]
\end{equation}
where $N[k]\sim\mathcal{N}(0,\sigma_N^2)$ is the complex Gaussian noise component at the receiving antenna, and without loss of generality, we assume unit variance, i.e., $\sigma_N^2=1$.
\subsection{Preliminaries}
In wireless communication systems, user energy efficiency quantitatively measures bits of information reliably transferred to a receiver per unit consumed energy at the transmitter. Basically, energy is consumed to power data processing circuitry and send the signal to the target destination through a wireless fading medium. In this paper, we consider system energy efficiency which is defined as
\begin{equation}
\eta=\frac{\text{Throughput}}{\text{Total consumed energy}}\,\,\,\,\,\,\,\,\,\,\,\,\,(\text{bits/Joule}).
\end{equation}
%An energy efficient strategy is essential in the presence of limited resources, and this can be a challenging task in MACs as there are more optimization variables. In the literature, few studies have focused on the EE of multiple access channels as stated in the previous section. However, there do not exist analytical characterizations of optimal transmitting policies and the corresponding system efficiency, to the best of our knowledge.\\
\indent We focus on the instantaneous achievable data rate per unit consumed energy, and consider only peak power constraint. Given the fading state realization, sum-rate capacity of a Gaussian MAC is
\begin{equation}
\label{Eq.Srate}
\mathcal{R}_{sum}[k]=T\log_2\Big(1+\sum_{i=1}^2\gamma_i[k]\Big)\,\,\,\,\,\,\,\,\,\,\,\,\,\,\,(\text{bits/Hz})
\end{equation}  
where $\gamma_i[k]=h_i[k]P_i[k]$. For simplicity, we eliminate the index $k$ for symbol duration in the sequel. Based on the above relation, sum-rate capacity is maximized when each user transmits at its peak power level. However, this might not be the optimal strategy when energy efficiency is considered, as will be discussed shortly.\\
\indent Let $P_{c_i}$ denote the circuit power consumption of user $i$, and assume that it is independent of the transmitted power level, $P_i$. Hence, the total energy expense of multiple users over an uplink operation interval of $T$ seconds becomes
\begin{equation}
\label{Eq.Etot}
E_{tot}=T\sum_{i=1}^2\Big(P_{c_i}+P_i\Big)=T\Big(P_c+\sum_{i=1}^2P_i\Big)\,\,\,\,\,\,\,\,\,\,\,\,(\text{Joules})
\end{equation} 
where $P_c=\sum_{i=1}^2P_{c_i}$. Thus, the instantaneous system energy efficiency over a block interval of T seconds is expressed  
\begin{equation}
\label{Eq.Eff}
\eta(P_i)=\frac{\log_2\Big(1+\sum_{i=1}^2\gamma_i\Big)}{P_c+\sum_{i=1}P_i}.
\end{equation}
\section{Energy efficiency with Separated ID and EH components}
Incorporating energy transfer along with information-bearing signal influences the optimal transmission policy and the way to define energy efficiency as well. Since spatial location of ID and EH components have direct impact on the analysis of energy efficiency and the corresponding optimal parameters, we consider separated scenarios where sources broadcast information and power simultaneously to spatially separated ID and EH components. Technically, EH component opportunistically harvests energy from the transmitted information-bearing signal which is intended for the ID component. This can improve the overall energy usage provided that the demand is satisfied with the energy-efficiency maximizing resource allocation. Meanwhile, further increase in demand for harvested energy requires sending the information-bearing RF signal at a higher power level. This in turn improves the data rate, but an improvement on the energy efficiency cannot be guaranteed.\\
\indent When the information-bearing signal is used to simultaneously energize the EH component, harvested energy should be subtracted from the total energy consumption in the formulation of the energy efficiency metric in order to reflect the net system energy consumption, as discussed in the literature \cite{Robert} \cite{Zhang}. Based on this remark, the system energy efficiency for multiple access channels with energy harvesting at EH is given as follows:
\begin{equation}
\label{Eq.eta_s}
\eta_s=\frac{\log_2\Big(1 + \sum_{i=1}^2 \gamma_i\Big)}{P_c+\sum_{i=1}^2P_i-\frac{\mathcal{E}_{hv}}{T}}
\end{equation}
where the numerator is the achievable sum-rate, denoted as $\mathcal{R}$, and the denominator indicates the net power consumed by the system. $\mathcal{E}_{hv}$ denotes harvested energy, and explicitly expressed as follows:
\begin{equation}
\begin{split}
\label{eq:R1}
%\mathcal{R}&= \log_2\Big(1 + \sum_{i=1} h_iP_i\Big)\,\,\,\,\,\,\,\, (\text{bps/Hz})\\
\mathcal{E}_{hv}&= \sum_{i=1}^2g_iP_i+\sigma_H^2\,\,\,\,\,\,\,\,\,\,\,\,\, (\text{Joules})
\end{split}
\end{equation}
where $g_i$ is the normalized channel power gain for the link between $i^{th}$ source and the EH component, and $\sigma_H^2$ is the variance of the noise component at the harvesting node. In the sequel, we assume that $T=1$. It is clear that both $\mathcal{R}$ and $\mathcal{E}_{hv}$ depend on $P_i$, and any increment in demand for harvested energy requires additional $\Delta P_i$ that could change the system efficiency by $\Delta\eta_s$. Hence, it is necessary to formulate optimization problems so as to trace the impact of harvested energy on the optimal operating points, and determine optimal transmit power as well as system energy efficiency. Thus, we have
\begin{equation}
\begin{split}
\label{Eq:P2}
\text{(P1):}\,\,\,\,\,\,\,\,\,&\max_{P_i\in\mathcal{P}}\,\,\,\eta_s\\
\text{subject to}\,\,\,\,\,\,\,\,&\mathcal{E}_{hv}\geq \chi
\end{split}
\end{equation}
where $\mathcal{P}=\bigcup_i\{P_i:0\leq P_i\leq P_i^{mx}\}$ and $\chi$ denotes the required harvested energy at the EH component. It is obvious that the sum-rate in (\ref{Eq.eta_s}) and total consumed energy in (\ref{eq:R1}) are concave and affine functions, respectively. Besides, both are differentiable. Hence, according to Proposition 2.9 stated in \cite{Alessio}, the system energy efficiency given in (\ref{Eq:P2}) satisfies the criteria for pseudo-concavity. This implies that KKT conditions are necessary and sufficient to obtain the globally optimal solution.\\ 
\indent Depending on the value of $\chi$, it is possible that opportunistically harvested energy can satisfy the constraint with strict inequality while information-bearing signal is transmitted at the energy-efficiency-maximizing power level. However, it would be fair to deduct  $\chi$ instead of $\mathcal{E}_{hv}$ from the total energy consumption for energy efficiency analysis assuming that the EH component harvests only the demand. Thus, problem P2 can be equivalently expressed as:
\begin{equation}
P_i=\arg\,\,\,\max_{P_i\in\mathcal{P}}\,\,\frac{\log_2\big(1 + \sum_{i=1}^2 \gamma_i\big)}{P_c+\sum_{i=1}^2P_i-\chi}. 
\end{equation}
Meanwhile, as the demand exceeds a certain threshold, it is necessary and sufficient to satisfy the constraint with equality. In such a case, we have 
\begin{equation}
P_i=\arg\,\,\,\max_{P_i\in\mathcal{P}}\,\,\eta_s\,\,\,\,\,\,\,\text{s.t.}\,\,\,\,\mathcal{E}_{hv}=\chi.  
\end{equation}
Based on the definition and property of pseudo-concavity, $\eta_s$ is a pseudo-concave function of $P_i$ under both conditions, given $\chi$. This guarantees that there is a unique optimal solution for P2, and the following proposition provides a closed-form analytical expression for the optimal transmission power level.
\begin{figure*}
	\begin{equation}
	\label{Eq.a}
	\begin{split}
	\frac{\partial \eta_s}{\partial P_i}=\frac{h_i}{\ln(2)\Big(1+\sum_i\gamma_i\Big)\Big(P_c+\sum_iP_i-\chi\Big)}-\frac{\log_2\Big(1+\sum_i\gamma_i\Big)}{\Big(P_c+\sum_iP_i-\chi\Big)^2}=0\\~\\
	\big(1+\sum_i\gamma_i\big)\ln\big(1+\sum_i\gamma_i\big)-\big(1+\sum_i\gamma_i\big)=h_i\big(P_c-\chi\big)-1+\sum_j (h_i-h_j)P_j
	\end{split}
	\end{equation}
\noindent\makebox[\linewidth]{\rule{18cm}{0.5pt}}
\end{figure*}
\begin{Prop}
	The optimally energy-efficient SWIPT power allocation strategy for two-user MACs with spatially separated ID and EH components is given by the following:
	\begin{subequations}
	\begin{align}		
	&\text{a})\,\,\text{If}\,\,0\,\,\leq\chi\leq\,\,\chi',\,\,\text{then}\nonumber\\
	\label{Eq.P_i}
	&\,\,\,\,\,\,\,P_i^*=\left\{
	\begin{array}{ll} \frac{e^{\mathcal{W}\big(\frac{h_i(P_c-\chi)-1}{e}\big)+1}-1}{h_i}&\hbox{$\chi\leq\chi^\ast$}\\ 
	\frac{\chi}{g_i}&\hbox{$\chi^\ast\leq\chi\leq\chi'$}\\
	\end{array}
	\right.\\
	&\,\,\,\,\,\,\,P_j^\ast=0\,;\,\,\,\,\,\,\,\,\,\,\,\,\,\,\,\,\,\,\,\,\,i,j\in\{1,2\}\\
	&\text{b})\,\,\text{If}\,\,\chi'\,\,\leq\chi\leq\,\,\chi^{mx},\,\,\text{then}\nonumber\\
	&\,\,\,\,\,\,\,P_i^*=\left\{
	\begin{array}{ll}
	P_i^{mx}&\hbox{$h_i>h_j, g_i>g_j$}\\
	\Big(\frac{g_je^{\mathcal{W}(\frac{A}{e})+1}-g_j-h_j\chi}{g_jh_i-g_ih_j},0\Big)^+&\hbox{$h_i>h_j, g_i<g_j$}\\
	\end{array}
	\right.\\
	\label{Eq.P_i-d}
	&\,\,\,\,\,\,\,P_j^\ast=\min\Big(\frac{\chi-g_iP_i^\ast}{g_j},P_j^{mx}\Big)
	\end{align}
	\end{subequations}
where $\chi$ is the required harvested energy and  $\chi^\ast$ is the harvested energy level at which the constraint starts being active. In addition, $\chi'$ and $\chi^{mx}$ denote the maximum energy that can be optimally harvested from user $i$ and the maximum achievable harvested energy from both users, respectively. Furthermore, $\mathcal{W}$ is the Lambert function, $A=a_{ij}P_c+\chi\Big(\frac{1-g_j}{g_j}a_{ij}-\frac{h_j}{g_j}\Big)-1$, $a_{ij}=\frac{h_ig_j-h_jg_i}{g_j-g_i}$.
\end{Prop}
\textit{Proof}: As noted earlier, the optimal strategy depends on the harvested energy demand, and hence we provide the details splitting the range of harvested energy into two levels where $\chi^{\ast}$ denotes a threshold for these. In the first case, we assume that the energy demand can be satisfied while maintaining maximum system energy efficiency, whereas in the second case, the energy constraint overrides the optimal system energy efficiency condition since the energy demand cannot be satisfied with the EE maximizing input.\\
(i) $0\,<\chi<\chi^{\ast}$: Here, the constraint can be satisfied with strict inequality, and the target is to determine the energy-efficiency maximizing solution. Since $\eta_s$ has a unique optimal value, this can be determined using the first order optimality condition given in (\ref{Eq.a}) at the top of next page. 
Let $\omega=1+\sum_{i=1}^2\gamma_i$ and $\Gamma=(h_i-h_j)P_j-1+h_i(P_c-\chi)$ for $i,j\in\{1,2\}$ with $i\neq j$. Thus, we have
\begin{equation}
\begin{split}
\omega\ln(\omega)-\omega&=\Gamma\\
e^{\ln(\frac{\omega}{e})}\ln(\frac{\omega}{e})&=\frac{\Gamma}{e}.\\
\end{split}
\end{equation}
After several mathematical manipulations, which are omitted due to space limitations, the solution to the above equation becomes
\begin{equation}
\label{Eq.omga}
	\omega^\ast=e^{\mathcal{W}(\frac{\Gamma}{e})+1}
\end{equation}
where $\mathcal{W}(\cdot)$ is the Lambert function. Then, substituting this solution into $\eta_s $ and taking the derivative, we have
\begin{equation}
\frac{\partial\eta_s}{\partial\Gamma}=-\Bigg[\frac{\mathcal{W}(\frac{\Gamma}{e})+1}{(\Gamma+e^{\mathcal{W}(\frac{\Gamma}{e})+1})^2}\Bigg].
\end{equation} 
This implies that energy efficiency is maximized at the minimum value of $\Gamma$, and this occurs when $P_j=0$ $\forall j\neq i$. Thus, the optimal transmit power for user $i$ is given as
\begin{equation}
\label{Eq.P_i2}
	P_i^\ast=\frac{e^{\mathcal{W}(\frac{\Gamma}{e})+1}-1}{h_i}\bigg|_{P_j=0}.
\end{equation}
The above expression is used to obtain the transmit power level until the demand reaches the threshold at which $\chi^\ast=g_iP_i^\ast$. This can be determined by substituting $P_i^\ast=\frac{\chi^\ast}{g_i}$ into (\ref{Eq.P_i2}) and solving for $\chi^\ast$. We obtain
\begin{equation}
W\Big(\frac{h_i(P_c-\chi^\ast)-1}{e}\Big)+1=\ln\Big(1+\frac{h_i}{g_i}\chi^\ast\Big).
\end{equation}
It is difficult to get a closed-form expression for $\chi^\ast$ from this equation, which, nevertheless, can be easily solved using numerical tools.\\
(ii)  $\chi^\ast\,<\chi<\chi^{mx}:$ Once the harvested energy demand exceeds this threshold, the constraint overrides the optimal energy efficiency condition. In such a case, we have 
\begin{equation}
\mathcal{L}= \eta_s + \mu (\mathcal{E}_{hv}-\chi)
\end{equation} 
and the corresponding KKT conditions are
\begin{subequations}
	\begin{align}
	\label{Eq.lag3}
	\frac{\partial\mathcal{L}}{\partial P_i}&=0\,\,\,\,\,\,\,\,\,\,\,\,\,\,\,\,\,\,\,
	\frac{\partial\mathcal{L}}{\partial \mu}=0\\
	\label{Eq.lag4}
	\mu\Big(\mathcal{E}_{hv}-\chi\Big)&=0\,\,\,\,\,\,\,\,\,\,\,\,\,\,\,\,
	\mu\geq 0,\,\,\,P_i\geq 0
	\end{align}
\end{subequations}
where $\mu$ is the Lagrange multiplier associated with the energy constraint. The optimality criteria (\ref{Eq.lag3}) is explicitly given in (\ref{Eq.d}) at the top of the next page. 
\begin{figure*}
	\begin{equation}
	\label{Eq.d}
	\begin{split}
	\frac{\partial \eta_s}{\partial P_i}=\frac{h_i}{\ln(2)\Big(1+\sum_{i=1}^2\gamma_i\Big)\Big(P_c+\sum_{i=1}^2P_i-\chi\Big)}-\frac{\log_2\Big(1+\sum_{i=1}^2\gamma_i\Big)}{\Big(P_c+\sum_{i=1}^2P_i-\chi\Big)^2}=\mu g_i\\
	\frac{\partial \eta_s}{\partial P_j}=\frac{h_j}{\ln(2)\Big(1+\sum_{i=1}^2\gamma_i\Big)\Big(P_c+\sum_{i=1}^2P_i-\chi\Big)}-\frac{\log_2\Big(1+\sum_{i=1}^2\gamma_i\Big)}{\Big(P_c+\sum_{i=1}^2P_i-\chi\Big)^2}=\mu g_j
	%\Big(1+\sum_ih_iP_i\Big)\ln\Big(1+\sum_ih_iP_i\Big)-\Big(1+\sum_ih_iP_i\Big)=h_i\big(P_c-\chi\big)-1+\sum_j (h_i-h_j)P_j
	\end{split}
	\end{equation}
\noindent\makebox[\linewidth]{\rule{18cm}{0.5pt}}
\end{figure*} 
According to these equations, there are two possible conditions that need to be considered independently.\\
(a) $h_i>h_j$ and $g_i>g_j$: Here, both channels between user $i$ and ID and EH components have better conditions than those associated with user $j$. In this case, (\ref{Eq.d}) holds true if either $P_i=0$ or $P_j=0$. However, for the same transmit power level, user $j$ achieves a lower EE. Thus, it is more efficient to have user $j$ to keep silent and allow user $i$ to transmit. The optimal power level becomes
\begin{equation}
P_i^\ast=\frac{\chi}{g_i}
\end{equation}
and the penalty coefficient can be determined by substituting this into (\ref{Eq.d}). The maximum energy that can be harvested from user $i$ only becomes
\begin{equation}
\chi'=g_iP_i^{mx}.
\end{equation}
(b) $h_i>h_j$ but $g_i<g_j$: In this case, it is possible to satisfy both equations with $P_i\neq 0$ and $P_j\neq 0$. Hence, applying 
\begin{equation}
	\frac{1}{g_i}\frac{\partial \eta_s}{\partial P_i}=\frac{1}{g_j}\frac{\partial \eta_s}{\partial P_j}\nonumber,
\end{equation}
we have
\begin{equation}
\label{Eq.hgij}
\bigg(\frac{h_i}{g_i}-\frac{h_j}{g_j}\bigg)\frac{h_j}{\big(1+\sum_i\gamma_i\big)}=\bigg(\frac{1}{g_i}-\frac{1}{g_j}\bigg)\frac{\ln\big(1+\sum_i\gamma_i\big)}{\big(P_c+\sum_iP_i-\chi\big)}.
\end{equation}
Meanwhile, the harvested energy constraint implies that $P_j=\frac{\chi-g_iP_i}{g_j}$, and substituting this into (\ref{Eq.hgij}), we get (\ref{Eq.hgij2}) given at the top of the next page. 
\begin{figure*}
\begin{equation}
\begin{split} 
\label{Eq.hgij2}
\frac{h_ig_j-h_jg_i}{g_j-g_i}\Big(P_c+P_i\big(1-\frac{g_i}{g_j}\big)+\chi\big(\frac{1}{g_j}-1\big)\Big)&=\bigg(1+\chi\frac{h_j}{g_j}+P_i\Big(h_i-\frac{h_jg_i}{g_j}\Big)\bigg)\ln\bigg(1+\chi\frac{h_j}{g_j}+P_i\Big(h_i-\frac{h_jg_i}{g_j}\Big)\bigg)\\
A&=z\ln(z)-z
\end{split} 
\end{equation}
where $A=P_c\Big(\frac{h_ig_j-h_jg_i}{g_j-g_i}\Big)+\chi\Big(\frac{1-g_j}{g_j}\big(\frac{h_ig_j-h_jg_i}{g_j-g_i}\big)-\frac{h_j}{g_j}\Big)-1$ and $z=1+\chi\frac{h_j}{g_j}+P_i\Big(h_i-\frac{h_jg_i}{g_j}\Big)$
\noindent\makebox[\linewidth]{\rule{18cm}{0.5pt}}
\end{figure*}
Hence, solving for $P_i$, we obtain
\begin{equation}
\label{Eq.Pij} 
P_i^\ast=\frac{g_je^{\mathcal{W}(\frac{A}{e})+1}-g_j-h_j\chi}{g_jh_i-g_ih_j}. 
\end{equation}
Accordingly, the maximum energy that can be harvested when $P_j^\ast=0$ can be obtained by setting $\chi'=g_iP_i^\ast$ in (\ref{Eq.Pij}). This can be determined solving 
\begin{equation}
\ln\Big(\frac{h_i}{g_i}\chi'+1\Big)=\mathcal{W}\left(\frac{A}{e}\right)\Big|_{\chi=\chi'}+1.
\end{equation}
This completes the proof.$\,\,\,\,\,\,\,\,\,\,\,\,\,\,\,\,\,\,\,\,\,\,\,\,\,\,\,\,\,\,\,\,\,\,\,\,\,\,\,\,\,\,\,\,\,\,\,\,\,\,\,\,\,\,\,\,\,\,\,\,\,\,\,\,\,\,\,\,\,\,\,\,\,\,\,\,\,\,\blacksquare$\\
\indent As can be observed from the proposition, the impact of harvested energy increment on the optimal transmitting power level depends on the region where it lies. For instance, in the region where $\chi<\chi^\ast$, $P_i^\ast$ decreases for each incremental $\Delta\chi$, assuming $h_i>h_j$, and the reason is as follows. First, it clear that $\Gamma$ decreases with increasing $\chi$, or vice versa. Knowing that Lambert function $\mathcal{W}(x)$ is a non-negative and increasing function for $x>0$, $e^{\mathcal{W}(\cdot)}$ is also an increasing function. Thus, incremental demand $\Delta\chi$ reduces $\mathcal{W}(\frac{\Gamma}{e})$, and of course $\omega^\ast$. Therefore, we conclude that optimal transmit power $P_i^\ast$ reduces further until $\chi=\chi^\ast$ based on (\ref{Eq.P_i}). However, once the demand exceeds $\chi^\ast$, the harvested energy constraint overrides the energy-efficiency-maximizing solution. As a result, the transmitted power level from user $i$ linearly increases with $\chi$ while user $j$ is still silent. The instant at which user $j$ becomes active depends on the channel power gain of the link between each user with ID and EH components. For instance, it is more energy efficient to only activate user $i$ until it reaches its peak provided $h_i>h_j$ and $g_i>g_j$. However, it is more beneficial to allocate resources for both users before user $i$ hits its peak when $g_j>g_i$. This implies that better energy efficiency is attained by introducing user $j$ instead of increasing the transmitted power level of user $i$ by an equivalent amount. This is because the linear gain $(g_j-g_i)\Delta P$ due to the incremental power, $\Delta P$, from user $j$ instead of user $i$ becomes significant compared to the logarithmic loss $\log\big((h_j-h_i)\Delta P\big)$. As a result, the energy efficiency becomes significant despite the reduction in the rate as $h_j<h_i$.
\begin{Cor}
	The optimal system energy efficiency given the harvested energy demand under spatially separated ID and EH architecture is
	\begin{equation}
	\eta_s^\ast(\chi)=\left\{
	\begin{array}{ll}
	\frac{h_i\log_2(\omega^\ast)}{\Gamma+\omega^\ast} & \hbox{$\chi<\chi^\ast$}\\~\\
	%\frac{\log_2(1+a_i\chi)}{P_c+b_i\chi} & \hbox{$;\chi^\ast<\chi<g_iP_i^{mx}$}\\~\\
	\frac{\log_2\big(1+(h_i-a_jg_i)P_i+a_j\chi\big)}{P_c+\big(1-\frac{g_i}{g_j}\big)P_i+b_j\chi} & \hbox{$\chi^\ast<\chi<\chi^{mx}$}
	\end{array}
	\right.
	\end{equation}
	where $\omega^\ast$ is as given in (\ref{Eq.omga}), $a_j=\frac{h_j}{g_j}$, and $b_j=\frac{1}{g_j}-1$.
\end{Cor}
\textit{Proof}: The optimal energy efficiency can be easily obtained by substituting the expression for $P_i^\ast$ and $P_j^\ast$ given in (\ref{Eq.P_i})-(\ref{Eq.P_i-d}) into (\ref{Eq.eta_s}).$\,\,\,\,\,\,\,\,\,\,\,\,\,\,\,\,\,\,\,\,\,\,\,\,\,\,\,\,\,\,\,\,\,\,\,\,\,\,\,\,\,\,\,\,\,\,\,\,\,\,\,\,\,\,\,\,\,\,\,\,\,\,\,\,\,\,\,\,\,\,\,\,\,\,\,\,\,\,\,\,\blacksquare$\\
\indent According to the above characterization, we observe that $\eta_s^\ast$ is always an increasing function with harvested energy demand for $\chi<\chi^\ast$. This is because the numerator decreases logarithmically with $\chi$ while the denominator reduces linearly. Hence, the ratio becomes an increasing function in $\chi$. On the other hand, $\eta_s^\ast$ is neither increasing nor decreasing function for $\chi>\chi^\ast$, and in such instances energy efficiency is generally expressed in $\frac{\log_2(ax+b)}{cx+d}$ form. This is clearly a pseudo-concave function, and there exists a point beyond which efficiency starts to decrease for each increment in $\chi$.
\section{Numerical Analysis}
\subsection{Simulation Model}
In this section, we provide numerical results for a network with two users and one access point where $h_1=0.8$, $h_2=0.4$, and $g_1=0.5$. For the link between user $2$ and EH receiver, we have considered two cases: $g_2=0.3$ and $g_2=0.8$ in order to investigate impact of channel conditions on optimal strategy. 
\subsection{System Energy Efficiency}
\begin{figure*}
	\begin{subfigure}{0.3\textwidth}
		\centering
		\includegraphics[width=0.9\textwidth]{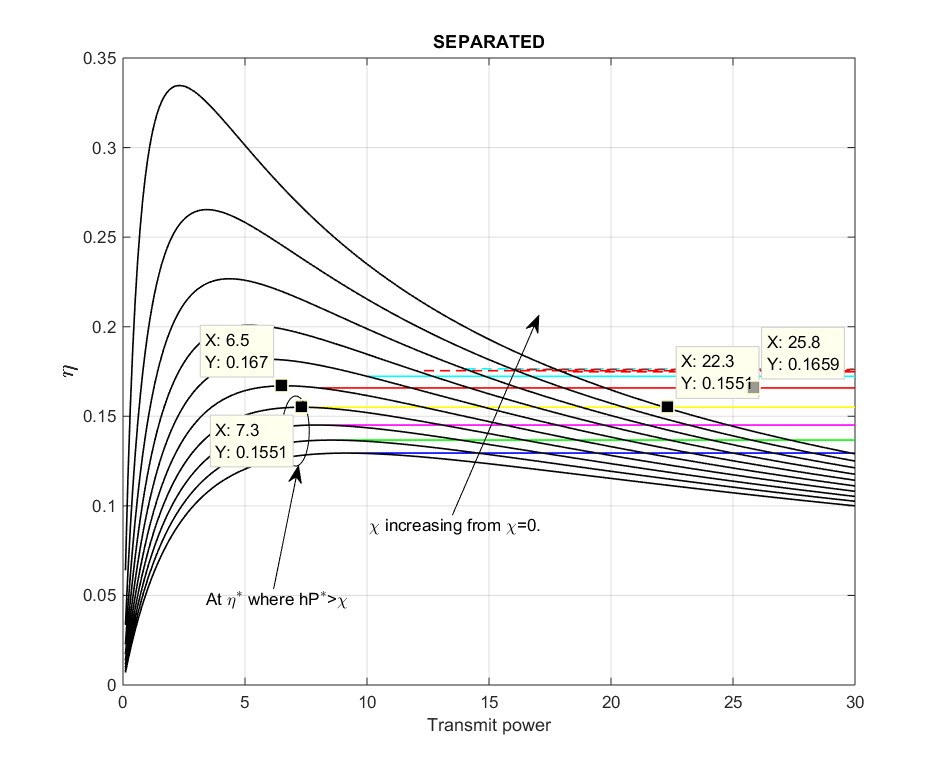}
		\caption{EE vs. transmitted power level}
		\label{Fig.eta-p}
	\end{subfigure}
	\begin{subfigure}{0.3\textwidth}
		\centering
		%  % Requires \usepackage{graphicx}
		\includegraphics[width=0.9\textwidth]{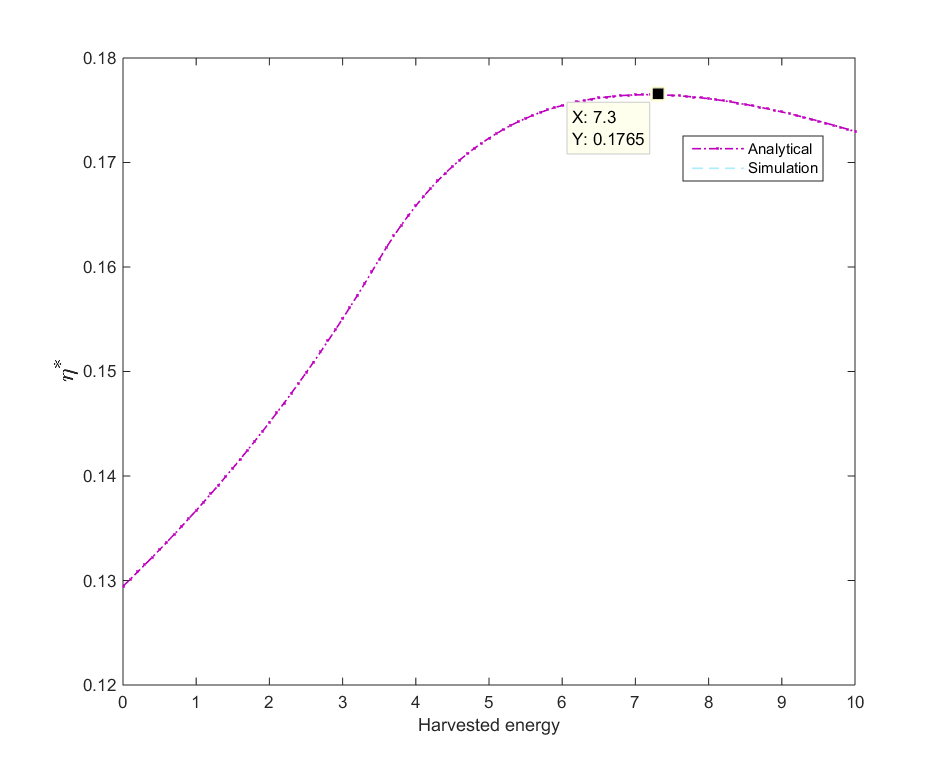}
		\caption{EE vs. $\chi$ in the single-user case}
		\label{Fig.eta-s}
	\end{subfigure}
	\begin{subfigure}{0.3\textwidth}
		\centering
		%  % Requires \usepackage{graphicx}
		\includegraphics[width=0.9\textwidth]{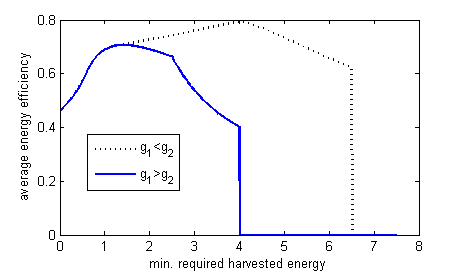}
		\caption{EE vs. $\chi$ in the two-user case}
		\label{Fig.eta_2U}
	\end{subfigure}
	\caption{Energy efficiency vs. $P_i$ and $\chi$}
	\label{Fig.eta_1U}
\noindent\makebox[\linewidth]{\rule{18cm}{0.5pt}}
\end{figure*}

\indent Fig. \ref{Fig.eta_1U} illustrates the characteristics of the optimal energy efficiency when wireless power transfer is incorporated along with data transmission. For instance, for separated ID and EH components, there is a possibility of satisfying the demand with the EE maximizing input as can be seen from Fig. \ref{Fig.eta-p}. However, further increase in harvested energy demand overrides efficiency optimality condition, i.e., the constraint forces the system to operate at a point below the most efficient condition. Furthermore, Fig. \ref{Fig.eta-s} clearly shows that the system's optimal energy efficiency has quasi-concave characteristics and for certain range of harvested energy demand, $\eta_s^\ast$ increases with $\chi$.
\begin{figure*}
	\centering
	\begin{subfigure}{.4\textwidth}
		\includegraphics[width=.85\linewidth]{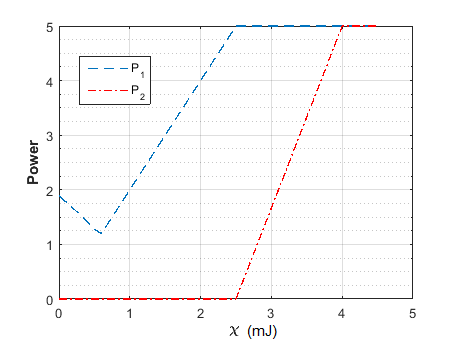}
		\caption{$P^\ast$ vs. $\chi$ given $g_i>g_j$}
		\label{Fig.P1}
	\end{subfigure}
	\begin{subfigure}{.4\textwidth}
		\centering
		\includegraphics[width=.85\linewidth]{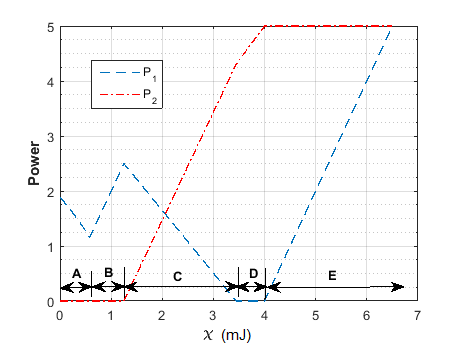}
		\caption{$P^\ast$ vs. $\chi$ given $g_i<g_j$}
		\label{Fig.P2}
	\end{subfigure}	
	\caption{Optimal transmit power vs. $\chi$ and $P_c$}
	\label{Fig.P_eta}	
\noindent\makebox[\linewidth]{\rule{18cm}{0.5pt}}
\end{figure*}
Similar characteristics are observed in the presence of multiple users based on Fig. \ref{Fig.eta_2U}. In the case of separated architecture, the overall performance depends on the corresponding channel gains of ID and EH components with each user. When $h_1>h_2$ and $g_1>g_2$, transmission from user $2$ always degrades the system efficiency. However, this is not the case if $g_1<g_2$ for a certain range of $\chi$, i.e., user $2$ could lead to better efficiency. 
\subsection{Transmit Power Level and Circuit Power}
\begin{figure}
		\centering
		\includegraphics[width=.7\linewidth]{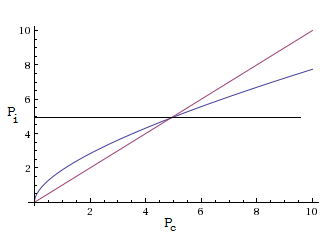}
		\caption{$P^\ast$ vs. $P_c$}
		\label{Fig.Pi-Pc}
\noindent\makebox[\linewidth]{\rule{9cm}{0.5pt}}		
\end{figure}

As can be seen from Fig.\ref{Fig.P_eta}, opportunistically harvesting energy under separated scenario reduces $P_1^\ast$ for certain ranges of $\chi$ while achieving better efficiency. Further increase in demand enforces signals to be sent at higher power levels, and this continues until the peak is reached at both users as shown in Fig.\ref{Fig.P1} for $g_1>g_2$. However, the optimal transmit power level is slightly modified when $h_1>h_2$ but $g_1<g_2$ as shown in Fig. \ref{Fig.P2}. In this case, $P_1$ first decreases and then increases with harvested energy demand as in regions A and B, respectively. However, it again starts to decrease in region C when additional energy obtained from user $2$ becomes significant compared with the corresponding reduction in information rate. This continues until the demand can not be satisfied by user $2$ as in region E, and each point beyond this guarantees that $P_2=P^{mx}$ and $P_1<P^{mx}$. Effect of circuit power consumption on the system energy efficiency is illustrated in Fig.\ref{Fig.Pi-Pc}. As can be seen, an increase in circuit power consumption shifts the optimal point of the characteristics below $P_1=P_c$ curve.
\section{Conclusion}
Energy efficiency is a key parameter demonstrating how efficient the system is utilizing available resources. In this paper, we have studied the performance of SWIPT in two-user multiple-access channels with energy efficiency as the major performance metric. In order to determine the optimal power control strategies maximizing the energy efficiency in the presence of energy harvesting constraints, we have formulated optimization problems for separated physical architectures. According to the analytical expressions, we have observed that optimal system energy efficiency has quasi-concave characteristics. Furthermore, harvested energy constraint can benefit the optimal energy efficiency. Besides, circuit power consumption triggers users to transmit signals at relatively lower power levels.

\end{document}